\documentclass{article}
\usepackage{amsmath}
\usepackage{amssymb}
\usepackage{amsfonts}
\usepackage{bm}
\usepackage[dvipdfmx]{graphicx}
\usepackage[dvipdfmx]{color}

\begin{document}

\title{
{\bf Dissipation in Langevin Equation
and Construction of Mobility Tensor
from Dissipative Heat Flow}}

\author{Takashi Uneyama\footnote{
E-mail: uneyama@mp.pse.nagoya-u.ac.jp}
\\
\\
JST-PRESTO, and 
Center for Computational Science, \\ Graduate School of Engineering, 
Nagoya University, \\
Furo-cho, Chikusa, Nagoya 464-8603, Japan
}

\date{}

\maketitle
%


\maketitle

\begin{abstract}
 The rheological behavior of a material is strongly related to the energy dissipation,
 and the understanding and modeling of dissipation is important from
 the view point of rheology. To study rheological properties with
 some mesoscopic and macroscopic dynamics models,
 the modeling of dynamic equation
 which appropriately incorporates the dissipation is important.
 Although there are several methods to
 construct mesoscopic and macroscopic dynamic equations, such as the Onsager's method,
 their validity {is} not fully clear. In this work, we theoretically analyze the dissipation
 in a mesoscopic Langevin equation in detail, from the view point of
 stochastic energetics. We show that the dissipative heat flow from the
 heat bath to the system plays an important role in the mesoscopic dynamics.
 The dissipative heat flow is unchanged under the variable transform, and
 thus it is a covariant quantity. We show that we can construct
 the Langevin equation and also perform a coarse-graining based on the
 dissipative heat flow. We can derive the mobility tensor from the
 dissipative heat flow, and construct the Langevin equation by combining
 it and the free energy. Our method can be applied to various systems,
 such as the dumbbell model and the diffusion type equation for the
 density field, to give the coarse-grained dynamic equations for them.
\end{abstract}



\section{Introduction}

To study the rheological behavior of some molecular and constitutive models, the energy
dissipation plays an important role. When we deform a viscoelastic material,
the material dissipates a part of the applied {work} into the heat.
The simple models such as the Maxwell and Voigt models have viscous
components (dash pots) which dissipate the energy.
In the linear viscoelasticity, the loss modulus $G''(\omega)$ can be related
to the energy dissipation under a small shear deformation at the angular frequency $\omega$\cite{Larson-book}. Without the
dissipation, we observe no viscous behavior and thus the system becomes totally elastic.
Therefore, the understanding
on the dissipation and the modeling which properly incorporates the dissipation
are important.
Also, the dissipation behavior depends on the time and length scales.
At the microscopic scale, we have basically no dissipation. 
At the mesoscopic and macroscopic scales, we observe dissipation but
the observed dissipation behavior depends on the scale (at least apparently).
As the scale increases, generally the contribution of the dissipation increases.
This is because the definition of the heat depends on the scale\cite{Sekimoto-book}.

To describe the dynamics in the mesoscopic systems,
the Langevin type equations \cite{vanKampen-book,Gardiner-book,Ottinger-book} are widely employed.
The Langevin equation 
naturally incorporates the dissipation as the frictional force.
In the macroscopic systems, phenomenological relaxation equations
such as the time-dependent Ginzburg-Landau (TDGL) equation are useful\cite{Onuki-book}.
The TDGL equation (without the noise term) is purely dissipative and the free energy of the system
monotonically decreases as the time evolves, which is a common property of relaxation processes.
The GENERIC (general equation for the nonequilibrium reversible-irreversible coupling) formalism\cite{Grmela-Ottinger-1997,Ottinger-Grmela-1997}
will be useful when there are both reversible and dissipative contributions in the system.
These equations are, however, not fully trivial from the view point of
fundamental statistical mechanics.

At the microscopic (or atomistic) scale, the system simply obeys classical mechanics.
The equations of motion can be expressed in several different ways.
A simple way is to employ the Newton's equation
of motion. However, the Lagrange's dynamic equation and Hamilton's canonical equation in the 
analytic mechanics are much more general and useful, although they may not be intuitive.
The Hamilton's principle states that in a mechanical system,
a path $\bm{q}(t)$ which minimizes the
action functional $\mathcal{S}[\bm{q}]$ is realized\cite{Landau-Lifshitz-book,Lanczos-book}:
\begin{equation}
 \label{hamilton_principle}
 \frac{\delta  \mathcal{S}[\bm{q}] }{\delta \bm{q}(t)} = 0 ,
\end{equation}
where $\delta / \delta \bm{q}(t)$ represents the functional derivative
with respect to $\bm{q}(t)$.
In this work, we call the condition in which the
functional derivative becomes zero as the {\em variational} principle, since
it is expressed in terms of the {\em variation} with respect to the path.
The Hamilton's principle and the Fermat's principle are variational principles
\cite{Lanczos-book}.
The action functional is expressed as the integral of
the Lagrangian:
\begin{equation}
 \label{action_lagrangian}
 \mathcal{S}[\bm{q}] = \int dt \, \mathcal{L}
  (\bm{q}(t),d\bm{q}(t)/dt) ,
\end{equation}
where $\mathcal{L}$ is the Lagrangian and is a function of $\bm{q}(t)$ and $d\bm{q}(t)/dt$.
Eqs~\eqref{hamilton_principle} and \eqref{action_lagrangian} give
the Euler-Lagrange equation and thus the Langrange's dynamic equation:
\begin{equation}
 \frac{\partial \mathcal{L}(\bm{q}(t),d\bm{q}(t)/dt)}{\partial \bm{q}(t)}
  - \frac{d}{dt} \frac{\partial \mathcal{L}(\bm{q}(t),d\bm{q}(t)/dt)}{\partial (d\bm{q}(t)/dt)} = 0.
\end{equation}
The Hamilton's principle \eqref{hamilton_principle} holds
even if we transform variable $\bm{q}(t)$ to another variable $\bm{q}'(t)$.
In this work, we may call a physical quantity or an equation is in a {\em covariant} form if
it does not change its form under the variable transform.
(Although the covariant property may look rather trivial, it is not
always satisfied. For example, as we will discuss later, the
free energy is not covariant.)

The Hamilton's principle \eqref{hamilton_principle} is powerful and useful because it is
variational and covariant. However, we should notice that
it can be usually applied
only to systems without any dissipation.
As we mentioned, for rheological properties
at the mesoscopic and macroscopic systems, the dissipation becomes essential.
Thus we should incorporate the dissipation into the dynamics, for example,
by introducing the frictional force. For such purposes, we usually employ
the Langevin equations or phenomenological TDGL type dynamic equations.
It is generally not clear whether the variational
and covariant properties are satisfied in these equations.
A general method which is variational and/or
covariant is desired, if it exists.

Many studies have been conducted to obtain a variational principle for
dissipative systems which corresponds
to the Hamilton's principle in the analytic mechanics\cite{Strutt-1873,Strutt-book,Bateman-1931,Riewe-1997,Lazo-Krumreich-2014,Galley-2013,Glavatskiy-2015,MartinezPerez-Ramirez-2018,Minguzzi-2015,Minguzzi-2015a,Allison-Pearce-Abbott-2014,Suzuki-2013}.
Rayleigh\cite{Strutt-1873,Strutt-book} proposed to use so-called the dissipation function to describe
Lagrange's dynamic equations with non-potential, friction force terms.
Onsager\cite{Onsager-1931,Onsager-1931a} also employed Rayleigh's dissipation function to describe the relaxation
process in general linear nonequilibrium systems.
However, both Rayleigh's and Onsager's methods are not in a variational form
(a functional derivative form).
Their dynamic equations do not have corresponding action functionals.
As we will show later, their methods should be interpreted rather as the minimization condition.

We can, nonetheless, construct the action for such dissipative systems.
Bateman\cite{Bateman-1931} proposed to couple two systems to construct the action functional.
In the Bateman's method, the dissipative system in which the total energy
monotonically decreases, is coupled to the dual system in which the total
energy monotonically increases. As a result, we obtain two set of dynamic
equations and one of them is for the dissipative system.
Bateman\cite{Bateman-1931} also proposed to use an explicitly time-dependent Lagrangian, as a
variant of the coupled systems method. However, these methods are not physically natural.
Riewe\cite{Riewe-1997} proposed to employ the fractional derivatives to express the dissipative
contributions. Although the Lagrangian with the fractional derivatives is
a direct generalization of that in the analytic mechanics,
it is questionable whether the the fractional derivatives have 
a physically reasonable interpretation or not. In addition, the fractional derivatives
are very sensitive to the initial and final conditions, which makes the
handling of the dynamic equation difficult.
Onsager and Machlup\cite{Onsager-Machlup-1953} derived the action functional which gives the probability of the
fluctuation around the most probable pathway. Although their action functional
can be utilized to obtain the information on the dynamics of dissipative
systems, it is not the generalization of the action integral in the
analytic mechanics. The Onsager-Machlup action should be interpreted as
the statistical weight factor for the Brownian motion.
Anyway, these methods are formal and mathematical, rather than
intuitive and physical. As Bauer\cite{Bauer-1931} showed, it is impossible to construct a variational
principle for dissipative systems, as a direct and natural generalization of the
Hamilton's principle~\eqref{hamilton_principle}.

Some methods and equations are still useful even if they are not in a variational form.
Recently Doi\cite{Doi-book,Doi-Zhou-Di-Xu-2019} demonstrated that some dynamic equations for soft matters can be
obtained by the Onsager's method. It would be informative to critically
and carefully examine the Onsager's method from the view point of modern statistical
physics. In the nonequilibrium statistical physics, even without a variational
principle, we can systematically describe the dynamics near equilibrium\cite{Itami-Sasa-2017}.
If the target system is not far from equilibrium,
the dynamics of the system can be described by the equilibrium free energy
(the thermodynamic potential) and the mobility tensor.
Such a system is often called the linear nonequilibrium system,
and we can apply the linear response theory to calculate the mobility\cite{Evans-Morris-book}.
The relation between this standard method and the Onsager's method is not
fully clear. In addition, as we stated, whether a method is covariant or not is not trivial.
We should carefully examine the covariant property of a method.

In this work, we apply the stochastic energetics\cite{Sekimoto-book} to the dynamic equation
in linear nonequilibrium systems, and derive a balance equation for the
free energy change rate and the heat flow from the bath to the system.
We consider a general mesoscopic system of which time evolution is given
as the Langevin equation.
Especially, we show that the dissipative part of the heat flow from the bath to the system plays
an important role, and it can be related to Rayleigh's 
dissipation function.
We discuss how we should interpret the Onsager's method, and then
we show that we can construct the dynamic equation in an alternative way.
The dissipative heat flow is in a covariant form and can be used to
derive the mobility tensor.
We show that we can systematically construct an equation {in a way} which is similar to
but different from the Onsager's method.
We also show a simple coarse-graining procedure for linear nonequilibrium systems,
based on the dissipative heat flow.
Finally, we apply our general method to some simple systems, as examples.
The mesoscopic models such as the dumbbell model and the interacting
Brownian particles can be consistently handled by our method.

\section{Model and Results}
\label{model_and_results}

\subsection{Dynamics Model}

We start from a general dynamics model for a linear nonequilibrium system,
where we do not have reversible currents nor memory kernels.
The time evolution of the system can be described reasonably by the
Fokker-Planck equation\cite{Risken-book}:
\begin{equation}
 \label{fokker_planck_equation}
 \frac{\partial P(\bm{x},t)}{\partial t}
  = \frac{\partial}{\partial x_{i}}
  \left[
   L_{ij}(\bm{x}) 
   \left[
    \frac{\partial \mathcal{F}(\bm{x})}{\partial x_{j}}
    P(\bm{x},t)
    + k_{B} T \frac{\partial P(\bm{x},t)}{\partial {x_{j}}}
   \right]
  \right] ,
\end{equation}
where $\bm{x} = \lbrace x_{i} \rbrace$ with $x_{i}$ being the $i$-th slow variable ($i = 1,2,\dots,n$),
$L_{ij}$ is the mobility tensor (the Onsager coefficient),
$\mathcal{F}(\bm{x})$ is the free energy of the system,
$k_{B}$ is the Boltzmann coefficient, and $T$ is the temperature.
The mobility tensor can depend on $\bm{x}$, but we require that it does not explicitly depend on time $t$.
(If the mobility explicitly depends on time, the analyses become quite complicated
\cite{Jeon-Chechkin-Metzler-2014,Uneyama-Miyaguchi-Akimoto-2015,Uneyama-Miyaguchi-Akimoto-2019}.)
We assume that the Einstein convention is applied to repeated indices (the summations are taken for $i$ and $j$ in eq~\eqref{fokker_planck_equation}).
The free energy can calculated from the partial partition function for
the microscopic degrees of freedom:
\begin{equation}
 \label{free_energy_from_partition_function}
 e^{-\mathcal{F}(\bm{x}) / k_{B} T}
  = \int d\bm{\Gamma} \, \delta(\bm{x} - \bm{X}(\bm{\Gamma})) e^{-\mathcal{H}(\bm{\Gamma}) / k_{B} T},
\end{equation}
where $\bm{\Gamma}$ represents the microscopic degrees of freedom,
$\mathcal{H}(\bm{\Gamma})$ is the Hamiltonian, and $\bm{X}(\bm{\Gamma})$ is an
expression of the slow variable as a function of the microscopic variable.
From the Onsager's reciprocal theorem\cite{Onsager-1931,Onsager-1931a,vanKampen-book,Doi-book}, $L_{ij}$ is symmetric, $L_{ij} = {L_{ji}}$.

We rewrite the Fokker-Planck equation \eqref{fokker_planck_equation} into
the Stratonovich type Langevin equation.
(The formalism of stochastic energetics requires the Langevin equation to
be the Stratonovich type\cite{Sekimoto-book}. The Ito type Langevin equation should be first
rewritten as the Stratonovich type, in prior to the analyses in the followings.)
We express the fluctuating
slow variable as $\bm{X}(t) = \lbrace X_{i}(t) \rbrace$ ($i = 1,2,\dots,n$).
From eq~\eqref{fokker_planck_equation}, we have the Stratonovich type 
Langevin equation as
\begin{equation}
  \label{langevin_equation}
  \frac{d  X_{i}(t)}{d t}
   = - L_{ij}(\bm{X}) 
    \frac{\partial \mathcal{F}(\bm{X})}{\partial X_{j}}
    + k_{B} T B_{ik}(\bm{X}) \frac{\partial B_{jk}(\bm{X})}{\partial X_{j}}
    + \sqrt{2 k_{B} T} B_{ij}(\bm{X}) w_{j}^{(\mathrm{S})}(t),
\end{equation}
where $B_{ij}(\bm{X})$ is the noise coefficient tensor
which satisfies $B_{ik}(\bm{X}) B_{jk}(\bm{X})
= L_{ij}(\bm{X})$,
and $w_{i}^{(\mathrm{S})}(t)$ is the Gaussian white noise. We express
the Stratonovich type stochastic term by the superscript ``(S)``.
The noise $w_{i}^{(\mathrm{S})}(t)$
should satisfy the fluctuation-dissipation relation:
\begin{equation}
 \langle w_{i}^{(\mathrm{S})}(t) \rangle = 0, \qquad
  \langle w_{i}^{(\mathrm{S})}(t) w_{j}^{(\mathrm{S})}(t') \rangle = \delta_{ij} \delta(t - t') .
\end{equation}
For some analyses and numerical calculations, the Ito type Langevin equation
is preferred. The Ito type Langevin equation for eq~\eqref{langevin_equation}
becomes\cite{Gardiner-book,Itami-Sasa-2017}
\begin{equation}
  \label{langevin_equation_ito}
  \frac{d  X_{i}(t)}{d t}
   = - L_{ij}(\bm{X}) 
    \frac{\partial \mathcal{F}(\bm{X})}{\partial X_{j}}
    + k_{B} T \frac{\partial L_{ij}(\bm{X})}{\partial X_{j}}
    + \sqrt{2 k_{B} T} B_{ij}(\bm{X}) w_{j}^{(\mathrm{I})}(t),
\end{equation}
where $w_{i}^{(\mathrm{I})}(t)$ is the Gaussian white noise, and
we use the superscript ``(I)'' to express the Ito type stochastic term.
The noise $w_{i}^{(\mathrm{I})}(t)$ satisfies the same fluctuation-dissipation relation {as one} for
the Stratonovich type Langevin equation: $\langle w_{i}^{(\mathrm{I})}(t)\rangle = 0$
and $\langle w_{i}^{\mathrm{(I)}}(t) w_{j}^{\mathrm{(I)}}(t') \rangle = \delta_{ij} \delta(t - t')$.

The fluctuation can be negligibly small for macroscopic systems.
For such a case, we may simply drop the noise term and the associated drift term in the Langevin
equation. The resulting dynamic equation
is deterministic and common for both the Stratonovich and Ito type equations:
\begin{equation}
  \label{relaxation_equation_from_langevin_equation}
  \frac{d  X_{i}(t)}{d t}
   = - L_{ij}(\bm{X}) 
    \frac{\partial \mathcal{F}(\bm{X})}{\partial X_{j}}.
\end{equation}
In this work, we are rather interested in mesoscopic systems where we
observe the thermal fluctuations.
In what follows, therefore, we mainly focus on the mesoscopic systems and do not drop
the noise term.

\subsection{Energy Balance}

Following Sekimoto\cite{Sekimoto-2007,Sekimoto-book}, we consider the energetics of this system.
The energy balance equation for the Langevin equation~\eqref{langevin_equation}
can be calculated rather straightforwardly. The change of the free energy,
$d\mathcal{F}$, is calculated as
\begin{equation}
 \begin{split}
  \label{free_energy_change}
  d\mathcal{F}
  & = \frac{\partial \mathcal{F}(\bm{X})}{\partial X_{i}}
  d X_{i} \\
  & = - 
   \left[  \frac{d  X_{i}(t)}{d t}
    - k_{B} T B_{ik}(\bm{X}) \frac{\partial B_{lk}(\bm{X})}{\partial X_{l}}
    - \sqrt{2 k_{B} T} B_{ik}(\bm{X}) w_{k}^{(\mathrm{S})}(t)  \right] 
  L^{-1}_{ij}(\bm{X})
  d X_{j} .
 \end{split}  
\end{equation}
The right hand side can be interpreted as the microscopic heat,
$d'\mathcal{Q}$, which
flows from the heat bath (or the environment) to the system.
(We use symbol $d'\mathcal{Q}$ instead of $d\mathcal{Q}$ since this
quantity is not an exact differential form.)
Thus we have
\begin{equation}
 \label{energy_balance_differential}
 d\mathcal{F} = d'\mathcal{Q} 
  = d'\mathcal{Q}_{d} + d'\mathcal{Q}_{f} ,
\end{equation}
with
\begin{equation}
 \label{dissipative_heat_change}
   d'\mathcal{Q}_{d}
  =  - \frac{dX_{i}}{dt} L^{-1}_{ij}(\bm{X}) d{X}_{j}, 
\end{equation}
\begin{equation}
 \label{fluctuating_heat_change}
  d'\mathcal{Q}_{f}
  = 
  \left[
  k_{B} T \frac{\partial B_{lk}(\bm{X})}{\partial X_{l}}
 + \sqrt{2 k_{B} T} w_{k}^{(\mathrm{S})}(t)  \right] 
 B^{-1}_{kj}(\bm{X}) dX_{j} . 
\end{equation}
We have decomposed the microscopic heat $d'\mathcal{Q}$ into two components,
$d'\mathcal{Q}_{d}$ and $d'\mathcal{Q}_{f}$.
$d'\mathcal{Q}_{d}$ and $d'\mathcal{Q}_{f}$ represent the dissipative
and fluctuating parts of the microscopic heat, respectively.
The dissipative heat $d'\mathcal{Q}_{d}$ is common for the mesoscopic and
macroscopic systems, but the fluctuating heat $d'\mathcal{Q}_{f}$ is absence
in the macroscopic system ($d'\mathcal{Q}_{f} = 0$). This is because the fluctuating heat
originates from the noise term in the Langevin equation.

In some cases, the change rate of the free energy
(the time derivative of the free energy) would be convenient than the
free energy change itself.
Thus we consider the rate of change of the free energy, $d\mathcal{F}/dt$.
Then we have
\begin{equation}
 \begin{split}
  \label{free_energy_change_rate}
  \frac{d\mathcal{F}(\bm{X}(t))}{dt} 
  & = 
 \frac{d'\mathcal{Q}\left(\bm{X}(t),{d\bm{X}(t)}/{dt},\bm{w}(t)\right)}{dt} \\
  & = \dot{\mathcal{Q}}_{d}
  \left(\bm{X}(t),\frac{d\bm{X}(t)}{dt} \right)
 + \dot{\mathcal{Q}}_{f}
 \left(\bm{X}(t),\frac{d\bm{X}(t)}{dt},\bm{w}(t)\right),
 \end{split}
\end{equation}
with
\begin{equation}
 \label{dissipative_heat_flow}
   \dot{\mathcal{Q}}_{d}(\bm{X},\dot{\bm{X}}) 
  =  - \dot{X}_{i} L^{-1}_{ij}(\bm{X}) \dot{X}_{j}, 
\end{equation}
\begin{equation}
 \label{fluctuating_heat_flow}
  \dot{\mathcal{Q}}_{f}(\bm{X},\dot{\bm{X}},\bm{w}) 
  = 
  \left[
  k_{B} T \frac{\partial B_{ik}(\bm{X})}{\partial X_{i}}
 + \sqrt{2 k_{B} T} w_{k}^{(\mathrm{S})}  \right] 
 B^{-1}_{kj}(\bm{X}) \dot{X}_{j}. 
\end{equation}
Here, $\dot{\mathcal{Q}}_{d}(\bm{X},\dot{\bm{X}}) $ and
$\dot{\mathcal{Q}}_{f}(\bm{X},\dot{\bm{X}},\bm{w}) $ correspond to
the dissipative and fluctuating heat flows from the bath to the system, respectively, for
given $\bm{X}$, $\dot{\bm{X}}$, and $\bm{w}$.
The positive definiteness of the mobility tensor $L_{ij}$ ensures the negativity of
the dissipative heat flow, $\dot{\mathcal{Q}}_{d}(\bm{X},\dot{\bm{X}}) \le 0$, for any $\bm{X}$ and $\dot{\bm{X}}$.
The energy balance equation \eqref{energy_balance_differential} can be rewritten
as
\begin{equation}
 \label{energy_balance_rate}
 \frac{\partial \mathcal{F}(\bm{X}(t))}{\partial X_{i}(t)} \dot{X}_{i}(t)
 = \dot{\mathcal{Q}}(\bm{X}(t),\dot{\bm{X}}(t))
 = \dot{\mathcal{Q}}_{d}(\bm{X}(t),\dot{\bm{X}}(t)) + \dot{\mathcal{Q}}_{f}(\bm{X}(t),\dot{\bm{X}}(t)) ,
\end{equation}
where $\dot{\mathcal{Q}}(\bm{X},\dot{\bm{X}})$ represents the total heat flow and
$\dot{\bm{X}}(t) = d\bm{X}(t)/dt$.

In some literature, Rayleigh's dissipation function is employed
to express the dissipation rate of the energy\cite{Strutt-1873,Strutt-book,Landau-Lifshitz-book,Doi-book}.
The Reyleigh dissipation function is defined as a function of
$\bm{X}$ and $\dot{\bm{X}}$ as
\begin{equation}
 \label{rayleigh_dissipation_function}
 \Psi(\bm{X},\dot{\bm{X}}) \equiv \frac{1}{2} \dot{X}_{i} L^{-1}_{ij}(\bm{X}) \dot{X}_{j}, 
\end{equation}
Comparing eqs~\eqref{dissipative_heat_flow} and \eqref{rayleigh_dissipation_function},
we find 
\begin{equation}
 \label{dissipative_heat_flow_and_rayleigh_dissipation_function}
 \dot{\mathcal{Q}}_{d}(\bm{X},\dot{\bm{X}}) = - 2 \Psi(\bm{X},\dot{\bm{X}}) 
\end{equation}
This means that, the Rayleigh dissipation function represents the {\em half}
of the heat flow from the system to the bath, in absence of the fluctuating
contribution. This is consistent with the standard interpretation of the
Rayleigh dissipation function\cite{Landau-Lifshitz-book}. The numerical factor $2$
in eq \eqref{dissipative_heat_flow_and_rayleigh_dissipation_function} comes from the fact that
Rayleigh {\em integrated} the heat change {\em with respect to $\dot{X}_{i}$},
in the following way:
\begin{equation}
 \label{rayleigh_dissipation_function_integral}
 \Psi(\bm{X},\dot{\bm{X}})
  = \int_{0}^{\dot{X}_{i}} \dot{X}_{i} L_{ij}(\bm{X}) d\dot{X}_{j} ,
\end{equation}
so that its partial derivative gives the dissipative force, $- \partial \Psi(\bm{X},\dot{\bm{X}}) / \partial \dot{X}_{i} = - L_{ij} \dot{X}_{j}$ \cite{Strutt-1873,Strutt-book}.
Eq~\eqref{rayleigh_dissipation_function_integral} is in analogy to the integration
of the thermodynamic force $- \partial \mathcal{F}(\bm{X}) /\partial X_{i}$ with respect to the variable $X_{i}$, to obtain the thermodynamic potential $\mathcal{F}(\bm{X})$.
However, from eq~\eqref{dissipative_heat_change}
we know that the microscopic dissipative heat change $d'\mathcal{Q}_{d}$ is defined for the small
change of $\bm{X}$, not for the small change of $\dot{\bm{X}}$.
In addition, the dissipative heat change is not an exact differential form and
it seems not to be reasonable to consider an integral quantity like eq~\eqref{rayleigh_dissipation_function_integral}.
Thus the Rayleigh dissipation function should
be interpreted as a result of a mathematical trick, rather than
a physically meaningful quantity.

The Rayleigh dissipation function is utilized to
describe the dynamic equation (without fluctuations) and is sometimes called the ``variational'' form
or the Onsager's ``variational'' principle\cite{Doi-book}:
\begin{equation}
 \label{pseudo_variational_form}
 \left. \frac{\partial}{\partial \dot{X}_{i}}
  \left[ 
   \frac{\partial \mathcal{F}(\bm{X})}{\partial X_{j}} \dot{X}_{j}
   + \Psi(\bm{X},\dot{\bm{X}}) 
   \right] \right|_{\dot{\bm{X}} = d\bm{X}/dt} = 0.
\end{equation}
It should be noticed that $\dot{\bm{X}}$ is assumed to be independent
of $\bm{X}$, when we calculate the partial differential with respect to $\dot{\bm{X}}$ in eq~\eqref{pseudo_variational_form}.
Such an equation cannot be expressed in a variational form as the Hamilton's principle~\eqref{hamilton_principle}.
Thus, although eq~\eqref{pseudo_variational_form} is refereed as the ``variational'' form in some literature,
in reality it is the minimization condition with respect to $\dot{\bm{X}}$ rather than the variation principle for $\bm{X}(t)$.
As Bauer\cite{Bauer-1931} showed, the simple variational principle, which is a direct generalization of
the Hamilton's principle in the analytical mechanics, does not exist for such a
system, unless we introduce some mathematical tricks which cannot
be physically justified.
Since the Rayleigh dissipation function has no reasonable physical meaning,
eq~\eqref{pseudo_variational_form} should also be interpreted as just a formal mathematical expression.

We consider an alternative equation from the view point of the stochastic energetics.
From eq~\eqref{energy_balance_rate}, what we should use is
the following equation, {instead}:
\begin{equation}
 \label{energy_balance_form}
 \left.
 \left[ \frac{\partial \mathcal{F}(\bm{X})}{\partial X_{j}}  \dot{X}_{j}
   - \dot{\mathcal{Q}}(\bm{X},\dot{\bm{X}},\bm{w}) \right]
 \right|_{\dot{\bm{X}} = d\bm{X}/dt} = 0 .
\end{equation}
Eq~\eqref{energy_balance_form} is not a variational form, nor a minimization condition with respect to
$\dot{\bm{X}}$. As clear from its derivation, it is nothing but the energy
balance equation.
If we ignore the fluctuating heat flow
($\dot{\mathcal{Q}}_{f} = 0$ and thus $\dot{\mathcal{Q}} = \dot{\mathcal{Q}}_{d}$),
eq~\eqref{energy_balance_form} can be rewritten as
\begin{equation}
 \label{energy_balance_form_dissipative}
 \left.
 \left[ \frac{\partial \mathcal{F}(\bm{X})}{\partial X_{j}}
   + \dot{X}_{i} L^{-1}_{ij} \right]   \dot{X}_{j}
 \right|_{\dot{\bm{X}} = d\bm{X}/dt} = 0 .
\end{equation}
Eq~\eqref{energy_balance_form_dissipative} {\em accidentally} gives the same equation as eq~\eqref{pseudo_variational_form}.
This partly supports the use of the Rayleigh dissipation function and
the minimization condition with respect to $\dot{\bm{X}}$.

Here we seek a physically reasonable method to construct the dynamic equation.
A key point is that the dissipative heat flow \eqref{dissipative_heat_flow} is in the quadratic form in
$\dot{\bm{X}}$. Thus from the dissipative heat flow, we can calculate
the (inverse) mobility tensor.
The mobility tensor can be obtained if we express the dissipative heat flow
into the bilinear form in $\dot{\bm{X}}$, as eq \eqref{dissipative_heat_flow}.
Or, if the explicit form of the dissipative heat flow is given, the following form may be simple and convenient:
\begin{equation}
 \label{mobility_from_heat_flow}
 L_{ij}^{-1}(\bm{X}) =
  - \frac{1}{2} \frac{\partial^{2} \dot{\mathcal{Q}}_{d}(\bm{X},\dot{\bm{X}})}{\partial \dot{X}_{i} \partial \dot{X}_{j}}.
\end{equation}
One may consider that the mobility tensor $L_{ij}$ calculated by eq~\eqref{mobility_from_heat_flow}
can depend on $\dot{\bm{X}}$, if the heat flow depends on $\dot{\bm{X}}$ in a
complex way. If the mobility tensor depends on $\dot{\bm{X}}$, it is beyond
the linear nonequilibrium and we cannot construct the Langevin equation
which is justified only for linear nonequilibrium systems. Thus the mobility tensor should be independent of $\dot{\bm{X}}$.
Once the mobility tensor is obtained,
we can construct the Langevin equation~\eqref{langevin_equation}
from the mobility tensor $L_{ij}(\bm{X})$ and the free energy $\mathcal{F}(\bm{X})$. In other words, we can construct
the dynamic equation from two scalar functions, $\dot{\mathcal{Q}}_{d}(\bm{X},\dot{\bm{X}})$
and $\mathcal{F}(\bm{X})$.

\subsection{Variable Transform and Covariant Property}

Although the Rayleigh dissipation function is not physically sound,
it is a scalar quantity and thus eq~\eqref{pseudo_variational_form} seems
to be essentially independent of the specific choice of the variable $\bm{X}$
{(covariant)}.
We check whether the relations shown above hold under the variable transform
from $\bm{X}$ to $\bm{X}'$ or not. The basic laws of physics should be
independent of the specific choice of the variable { and thus covariant}
(which is especially
clear in the theory of relativity\cite{Pauli-book}).

The new slow variable $\bm{X}'$ can be
interpreted as a function of $\bm{X}$, as $\bm{X}'(\bm{X})$.
Conversely, the original slow variable $\bm{X}$ can be interpreted as $\bm{X}(\bm{X}')$, in a similar way.
We introduce the Jacobian matrix as
$J_{ij} \equiv {\partial X'_{i}}/{\partial X_{j}}$, and its inverse becomes
$J_{ij}^{-1} = {\partial X_{i}}/{\partial X_{j}'}$.
Then, from the Langevin equation~\eqref{langevin_equation}, we have
\begin{equation}
 \label{langevin_equation_transformed}
  \begin{split}
   \frac{d  X_{i}'(t)}{d t}
   & = J_{ik}
   \left[
   - L_{kj}(\bm{X}) 
    \frac{\partial \mathcal{F}(\bm{X})}{\partial X_{i}}
    + k_{B} T B_{kl}(\bm{X}) \frac{\partial B_{jl}(\bm{X})}{\partial X_{j}}
    + \sqrt{2 k_{B} T} B_{kj}(\bm{X}) w_{j}^{(\mathrm{S})}(t) \right]
   \\
   & = - L_{jl}'
   \frac{\partial \mathcal{F}'}{\partial X_{j}'}
    + k_{B} T B_{ik}' \frac{\partial B_{jk}'}{\partial X_{j}'}
    + \sqrt{2 k_{B} T} B_{jl}' w_{l}^{(\mathrm{S})}(t) 
  \end{split}
\end{equation}
where
\begin{equation}
 \label{free_energy_transformed}
 \mathcal{F}'(\bm{X}') = \mathcal{F}(\bm{X}(\bm{X}'))
  + k_{B} T \ln (\det \bm{J}) ,
\end{equation}
\begin{equation}
 \label{mobility_tensor_transformed}
 L_{ij}'(\bm{X}') = J_{ik} L_{kl} J_{lj}, \qquad
  B_{ij}'(\bm{X}') = J_{ik} B_{kj} .
\end{equation}
Thus we find that the Langevin equation~\eqref{langevin_equation} can 
be transformed into the same form by the variable transform. However,
it should be noticed that the free energy $\mathcal{F}(\bm{X})$ in
eq~\eqref{langevin_equation} is
generally different from $\mathcal{F'}(\bm{X}')$ in eq~\eqref{langevin_equation_transformed}.
That is, the free energy is not in a covariant form.
Although this may not look intuitive, it is consistent with the equilibrium
statistical mechanics. The free energy for $\bm{X}'$ should be defined as
\begin{equation}
 \label{free_energy_trnasform_equilibrium}
 e^{-\mathcal{F}'(\bm{x}') / k_{B} T}
  \equiv \int d\bm{x} \, \delta(\bm{x}' - \bm{X}'(\bm{x})) e^{-\mathcal{F}(\bm{x}) / k_{B} T} .
\end{equation}
As Nakamura\cite{Nakamura-2018} explicitly stated, the free energy $\mathcal{F}'$ generally depends on
the Jacobian matrix for the variable transform.
Eq~\eqref{free_energy_trnasform_equilibrium} can be modified as
\begin{equation}
 \label{free_energy_trnasform_equilibrium_modified}
  e^{-\mathcal{F}'(\bm{x}') / k_{B} T}
  = \frac{1}{\det \bm{J}} e^{-\mathcal{F}(\bm{X}(\bm{x}')) / k_{B} T} ,
\end{equation}
and this gives eq~\eqref{free_energy_transformed}.
Eqs~\eqref{free_energy_transformed} and \eqref{free_energy_trnasform_equilibrium_modified}
mean that the free energy is {\em not} in a covariant form. Thus, when we
transform the slow variable, we should carefully calculate the free energy
for the new slow variable.

The energy balance equation is
\begin{equation}
 \begin{split}
  \label{free_energy_change_rate_transformed}
  \frac{d\mathcal{F}'(\bm{X}'(t))}{dt} 
  & = \dot{\mathcal{Q}}'_{d}
  \left(\bm{X}'(t),\frac{d\bm{X}'(t)}{dt} \right)
 + \dot{\mathcal{Q}}'_{f}
 \left(\bm{X}'(t),\frac{d\bm{X}'(t)}{dt},\bm{w}(t)\right),
 \end{split}
\end{equation}
with
\begin{equation}
 \label{dissipative_heat_flow_transformed}
   \dot{\mathcal{Q}}_{d}'(\bm{X}',\dot{\bm{X}}') 
  =  - \dot{X}_{i}' {L'_{ij}}^{-1} (\bm{X}') \dot{X}_{j}', 
\end{equation}
\begin{equation}
 \label{fluctuating_heat_flow_transformed}
  \dot{\mathcal{Q}}_{f}'(\bm{X}',\dot{\bm{X}}',\bm{w}) 
  = 
  \left[
  k_{B} T \frac{\partial B_{ik}'(\bm{X}')}{\partial X_{k}'}
 + \sqrt{2 k_{B} T} w_{i}^{(\mathrm{S})}  \right] 
  {B'_{ij}}^{-1}(\bm{X}) \dot{X}_{j}'. 
\end{equation}
From eq~\eqref{free_energy_transformed}, the left hand side of
eq~\eqref{free_energy_change_rate_transformed} is {\em not} equal to that
of eq~\eqref{free_energy_change_rate}
($d\mathcal{F}/dt \neq d\mathcal{F}'/dt$), although the energy balance
equation itself holds. This is because the form of the free energy
(strictly speaking, the entropy term in the free energy) depends on the
choice of the variable. This can be understood that some terms in the 
right hand side in eq~\eqref{free_energy_change_rate} moved to the left
hand side in eq~\eqref{free_energy_change_rate_transformed},
under the variable transform.
However, the dissipative heat flow is unchanged under this transform:
\begin{equation}
 \label{dissipative_heat_flow_relation}
   \dot{\mathcal{Q}}_{d}'(\bm{X}',\dot{\bm{X}}') 
   = \dot{\mathcal{Q}}_{d}(\bm{X},\dot{\bm{X}})  .
\end{equation}

The dissipative heat flow $\dot{\mathcal{\mathcal{Q}}}_{d}$ is therefore independent of the specific choice of the variable
and thus is in a covariant form.
Therefore, the mobility tensor for $\bm{X}'$ can be directly obtained from
the dissipative heat flow for the original variable, in the same form as
eq~\eqref{mobility_from_heat_flow}:
\begin{equation}
 \label{mobility_from_heat_flow_transformed}
 {L_{ij}'}^{-1}(\bm{X}') 
  = - \frac{1}{2} \frac{\partial^{2} \dot{\mathcal{Q}}_{d}(\bm{X}',\dot{\bm{X}}')}{\partial \dot{X}_{i}' \partial \dot{X}_{j}'}.
\end{equation}
Thus we conclude that, without explicitly calculating the variable
transform for the Langevin equation, we can construct the Langevin
equation for a given form of the slow variable, only from eqs~\eqref{free_energy_transformed} and \eqref{mobility_from_heat_flow_transformed}.
A procedure to derive the Langevin equation for a given variable $\bm{X}$ can be
summarized as follows.
\begin{enumerate}
 \item Calculate the free energy $\mathcal{F}(\bm{X})$. This is independent
       of the details of dynamics, and can be calculated from the partial
       partition function as eq~\eqref{free_energy_from_partition_function}.
       (However, the free energy is not covariant and it should be calculated
       for this {specific} variable $\bm{X}$, not for other variables.)
 \item Calculate the dissipative heat flow $\dot{\mathcal{Q}}_{d}$. This
       quantity is covariant (independent of the choice of the variable),
       and thus any variable transforms can be used to calculate it.
 \item Calculate the mobility tensor $L_{ij}(\bm{X})$ from $\dot{\mathcal{Q}}_{d}$
       via eq~\eqref{mobility_from_heat_flow}, and then
       construct the Langevin equation from $\mathcal{F}(\bm{X})$ and $L_{ij}(\bm{X})$.
       Both the Stratonovich and the Ito types (eqs \eqref{langevin_equation}
       and \eqref{langevin_equation_ito}) can be employed.
\end{enumerate}
This is a general method to construct the mesoscopic dynamic equation, and
is one of the main results {in} this work. Following this method, we can utilize the covariant property
and construct the dynamic equation systematically. 

Here, it would be informative to mention about so-called the Rayleighian.
Eq~\eqref{pseudo_variational_form} is sometimes rewritten by introducing
the Rayleighian $\mathcal{R}$\cite{Doi-book}:
\begin{equation}
 \label{rayleighian_definition}
  \mathcal{R}(\bm{X},\dot{\bm{X}}) \equiv \frac{\partial \mathcal{F}(\bm{X})}{\partial X_{j}} \dot{X}_{j}
   + \Psi(\bm{X},\dot{\bm{X}}) 
   = \frac{\partial \mathcal{F}(\bm{X})}{\partial X_{j}} \dot{X}_{j}
   + \frac{1}{2} \dot{X}_{i} L_{ij}^{-1} \dot{X}_{j} ,
\end{equation}
\begin{equation}
 \label{pseudo_variational_form_rayleighian}
 \left. \frac{\partial \mathcal{R}(\bm{X},\dot{\bm{X}})}{\partial \dot{\bm{X}}}  \right|_{\dot{\bm{X}} = d\bm{X}/dt} = 0.
\end{equation}
As we showed, the dissipative heat flow $\dot{\mathcal{Q}}_{d}$
is in a covariant form (eq~\eqref{dissipative_heat_flow_relation}), and 
from eq~\eqref{dissipative_heat_flow_and_rayleigh_dissipation_function}, the Rayleigh dissipation function $\Psi$ is also in a covariant form.
However, the free energy $\mathcal{F}$ is {\em not} in a covariant form,
as clearly shown in eq~\eqref{free_energy_transformed}. Therefore,
the Rayleighian varies under the variable transform and $\mathcal{R}'(\bm{X}',\dot{\bm{X}}')
\neq \mathcal{R}(\bm{X},\dot{\bm{X}})$, in general.
As a result, the
minimization condition \eqref{pseudo_variational_form_rayleighian} is not
in a covariant form, neither. This means that, if we calculate the
Rayleighian for a specific variable $\bm{X}$ as $\mathcal{R}(\bm{X},\dot{\bm{X}})$
but minimize it with respect to other variable $\bm{X}'$ as
$\partial \mathcal{R}(\bm{X},\dot{\bm{X}}) / \partial \dot{\bm{X}}'|_{\dot{\bm{X}}' = d\bm{X}'/dt} = 0$, the resulting dynamic equation for $\bm{X}'$ is not correct
(unless the Jacobian matrix is independent of $\bm{X}'$).
The Onsager's ``variational'' method which utilizes the Rayleighian is, therefore,
not variational nor covariant. (Strictly speaking, the Rayleigh's original
method which utilizes the microscopic Lagrangian and the dissipation function is covariant,
since the Lagraigian is covariant. But of course, it is not variational.)

\subsection{Coarse-Graining}

The dissipative heat flow may be also utilized for the coarse-graining.
To obtain the coarse-grained dynamic equation, we statistically eliminate
relatively fast variables. Because the choice of the slow variable is
rather arbitrary, we set $\bm{X} = \lbrace \bm{Y}, \bm{\Theta}\rbrace$,
where $\bm{Y} = \lbrace Y_{i} \rbrace$ ($i = 1,2,\dots,m$) is the set of variables for the coarse-grained
system, and $\bm{\Theta} = \lbrace \Theta_{\alpha} \rbrace$ ($\alpha = m + 1,2,\dots,n$)
is the set of the relatively fast variables to be eliminated. In this subsection, we use subscripts $i,j,k,\dots$ for
the coarse-grained slow variables and subscripts $\alpha,\beta,\gamma,\dots$ for the relatively fast variables.

The free energy for the coarse-grained system $\bar{\mathcal{F}}(\bm{Y})$ is
\begin{equation}
 \label{free_energy_coarse_graining}
 e^{-\bar{\mathcal{F}}(\bm{Y}) / k_{B} T}
  = \int d\bm{\Theta} \,  e^{-\mathcal{F}(\bm{Y}, \bm{\Theta}) / k_{B} T} .
\end{equation}
The dissipative heat flow of the system can be expressed as
\begin{equation}
 \label{dissipative_heat_flow_coarse_graining}
 \begin{split}
  \dot{\mathcal{Q}}_{d} 
  & = 
  - \begin{bmatrix}
     \dot{Y_{i}} & \dot{\Theta}_{\alpha}
  \end{bmatrix}
  \begin{bmatrix}
   M_{i j} & M_{i \beta} \\
   M_{\alpha j} & M_{\alpha \beta}
  \end{bmatrix}
  \begin{bmatrix}
   \dot{Y_{j}} \\ \dot{\Theta}_{\beta}
  \end{bmatrix} \\
  & = - \dot{Y}_{i} M_{ij} \dot{Y}_{j}
  - 2 \dot{Y}_{i} M_{i\alpha} \dot{\Theta}_{\alpha}
  - \dot{\Theta}_{\alpha} M_{\alpha \beta} \dot{\Theta}_{\beta} ,
 \end{split}
\end{equation}
where we have defined the components of the inverse mobility tensor via
\begin{equation}
  \begin{bmatrix}
   M_{i k} & M_{i \gamma} \\
   M_{\alpha k} & M_{\alpha \gamma}
  \end{bmatrix}
  \begin{bmatrix}
   L_{k j} & L_{k \beta} \\
   L_{\gamma j}& L_{\gamma \beta}
  \end{bmatrix}
 =
 \begin{bmatrix}
  \delta_{ij} & 0 \\
  0 & \delta_{\alpha\beta}
 \end{bmatrix} .
\end{equation}
The dissipative heat flow is a function of $\bm{Y}$, $\bm{\Theta}$, $\dot{\bm{Y}}$, and
$\dot{\bm{\Theta}}$.
We want to eliminate the relatively fast variables $\bm{\Theta}$ and $\dot{\bm{\Theta}}$, to obtain the
effective dissipative heat flow for the coarse-grained slow variables.
For $\bm{\Theta}$, we can simply take the statistical average over the local equilibrium distribution:
\begin{equation}
 P_{\text{leq}}(\bm{\Theta}|\bm{Y}) = 
  e^{\bar{\mathcal{F}}(\bm{Y}) / k_{B} T
  - \mathcal{F}(\bm{Y},\bm{\Theta}) / k_{B} T}
\end{equation}
The equilibrium distribution of $\dot{\bm{\Theta}}$ is generally not that simple.
However, because we employed the local equilibrium distribution for $\bm{\Theta}$,
it would be natural to expect that $\dot{\bm{\Theta}}$ is characterized by a
simple fluctuation around the local equilibrium. It may depend on the
slow degrees of freedom, $\bm{Y}$ and $\dot{\bm{Y}}$. We can rewrite
the dissipative heat flow as follows:
\begin{equation}
 \label{dissipative_heat_flow_coarse_graining_modified}
  \dot{\mathcal{Q}}_{d} 
   = - \dot{Y}_{i}
  (M_{ij}
    - M_{i\alpha}
  M_{\alpha \beta}^{-1} M_{\beta j}
  ) \dot{Y}_{j}
  - (\dot{\Theta}_{\alpha} + M_{\alpha \gamma}^{-1} M_{\gamma i} \dot{Y}_{i})
  M_{\alpha \beta} (\dot{\Theta}_{\beta} + M_{\beta\delta}^{-1} M_{\delta j} \dot{Y}_{j}) .
\end{equation}
Here, we may interpret the second term in eq~\eqref{dissipative_heat_flow_coarse_graining_modified}
as the dissipative heat flow due to the relatively fast degrees of freedom.
Following this interpretation, we assume the following phenomenological
form for the local equilibrium distribution of $\dot{\bm{\Theta}}$:
\begin{equation}
 P_{\text{leq}}(\dot{\bm{\Theta}}|\dot{\bm{Y}},\bm{Y},\bm{\Theta})
  = \frac{1}{\sqrt{2 \pi \det \bm{C}}} 
  e^{- (\dot{\Theta}_{\alpha} + M^{-1}_{\alpha \gamma} M_{\gamma i} \dot{Y}_{i})
  G_{\alpha \beta}^{-1} (\dot{\Theta}_{\beta} + M^{-1}_{\beta \delta} M_{\delta j} \dot{Y}_{j}) / 2},
\end{equation}
where $G_{\alpha \beta}(\bm{Y},\bm{\Theta})$ is the covariance {matrix}, and 
can depend both on $\bm{Y}$ and $\bm{\Theta}$.

Then we have the following approximate form for the coarse-grained dissipative heat flow:
\begin{equation}
 \begin{split}
  \dot{\bar{\mathcal{Q}}}_{d}(\bm{Y},\dot{\bm{Y}}) 
  & \approx \int d\bm{\Theta} d\dot{\bm{\Theta}} \,
  \, P_{\text{leq}}(\bm{\Theta}|\bm{Y}) P_{\text{leq}}(\dot{\bm{\Theta}}|\dot{\bm{Y}},\bm{Y},\bm{\Theta}) 
  \dot{\mathcal{Q}}_{d}   \\
  & = \int d\bm{\Theta} \,
  \, P_{\text{leq}}(\bm{\Theta}|\bm{Y}) \left[ - \dot{Y}_{i}
  (M_{ij}
    - M_{i\alpha}
  M_{\alpha \beta}^{-1} M_{j\beta}
  ) \dot{Y}_{j}
  - G_{\alpha \beta}
  M_{\alpha \beta} \right] \\
  & = - \dot{Y}_{i} \bar{L}^{-1}_{ij} \dot{Y}_{j}
  + \dot{\bar{\mathcal{Q}}}_{d}^{(\text{leq})}(\bm{Y}) ,
 \end{split}
\end{equation}
with
\begin{equation}
 \label{coarse_grained_mobility_tensor}
 \bar{L}_{ij}^{-1} = \int d\bm{\Theta} \, P_{\text{leq}}(\bm{\Theta}|\bm{Y})
    (M_{ij}
    - M_{i\alpha}
  M_{\alpha \beta}^{-1} M_{\beta j}
  ) ,
\end{equation}
\begin{equation}
 \dot{\bar{\mathcal{Q}}}_{d}^{(\text{leq})}(\bm{Y}) = - \int d\bm{\Theta} \,
  \, P_{\text{leq}}(\bm{\Theta}|\bm{Y}) 
   G_{\alpha \beta} M_{\alpha \beta} .
\end{equation}
Here, $\dot{\bar{\mathcal{Q}}}_{d}^{(\text{leq})}$ is independent of $\dot{\bm{Y}}$ and does not
contribute to the mobility tensor. In addition, it should be balanced to
the statistically averaged fluctuating heat flow, otherwise the energy balance
equation is violated. Thus we may simply ignore $\dot{\bar{\mathcal{Q}}}_{d}^{(\text{leq})}$.
Further, if we assume that the coarse-grained fluctuating heat flow has the same form
as eq~\eqref{fluctuating_heat_flow}, we recover the energy balance equation which
has the same form as eq~\eqref{free_energy_change_rate}.

Eq~\eqref{coarse_grained_mobility_tensor} gives the coarse-grained mobility tensor.
By calculating the mobility tensor for $\bm{Y}$ by inverting 
eq~\eqref{coarse_grained_mobility_tensor},
and introducing the noise tensor $\bar{B}_{ij}$ via $\bar{B}_{ik} \bar{B}_{jk} = \bar{L}_{ik}$,
finally we can construct the approximate Langevin equation for
the coarse-grained variable:
\begin{equation}
 \label{coarse_grained_langevin_equation}
  \frac{dY_{i}(t)}{dt} 
    =
  - \bar{L}_{ij} \frac{\partial \bar{\mathcal{F}}(\bm{Y})}{\partial Y_{j}}
  + k_{B} T \bar{B}_{ik}(\bm{Y}) \frac{\partial \bar{B}_{jk}(\bm{Y})}{\partial Y_{j}}
  + \sqrt{2 k_{B} T} \bar{B}_{ij}(\bm{Y}) w_{j}^{(\mathrm{S})} .
\end{equation}
Of course, the obtained Langevin equation
\eqref{coarse_grained_langevin_equation} by our method is not exact.
In a standard coarse-graining procedure, the elimination of relatively fast variables should be
conducted with the projection operator method, and the result becomes
the generalized Langevin equation with the memory kernel\cite{Kawasaki-1973}.
However, the memory kernel is difficult to handle with, and the
memoryless Langevin equation under the Markovian approximation is
widely employed as an approximate dynamic equation.
(A recent work\cite{Uneyama-Nakai-Masubuchi-2019} showed that the generalized
Langevin equation can be approximated by a Langevin equation with an effective
mobility tensor, if the memory effect is weak.)
Our derivation
of the coarse-grained Langevin equation does not use the projection
operator and simple (although the validity and accuracy of approximations
are not fully justified). Due to its simplicity, it will be useful
for some applications.
This result is also one of the main results {in} this work.
Both the construction of the Langevin equation and the coarse-graining
can be performed in a similar way, by utilizing the dissipative
heat flow.

%


\section{Examples}
\label{examples}

In Sec~\ref{model_and_results}, we obtained a general method to construct
the Langevin equation {for} the system, and also a general coarse-graining method.
These methods are rather formal and it is not clear whether they can be really applied to
systems which are interesting from the rheological viewpoint.
In this section, we apply our method to some simple systems, as examples,
and study how our methods work.

\subsection{Dumbbell model}

As a simple example, we consider the dynamics of a dumbbell model\cite{Ottinger-book}.
We consider a dumbbell which consists of two particles connected by a
bond potential, and express the particle positions as $\bm{r}_{1}$ and $\bm{r}_{2}$,
and the bond potential as $U(\bm{r}_{1} - \bm{r}_{2})$. This bond potential
can be interpreted as the free energy of the dumbbell.
Also, we express the mobility tensor as $L_{ij}(\bm{r}_{1},\bm{r}_{2})$ ($i, j = 1, 2$).
The mobility tensor can depend on $\bm{r}_{i}$ in a complex way.
For example, if the friction coefficient is position-dependent, the elements
{ of the mobility tensor} depend on the positions.
The motion of the solvent around the dumbbell will be
(approximately) expressed as off-diagonal elements.
(Strictly speaking, the mobility tensor is generally not isotropic.
In this work, for simplicity, we limit ourselves to an isotropic mobility tensor model.)
Then, the Langevin equation becomes
\begin{equation}
 \frac{d\bm{r}_{i}(t)}{dt}
  = - L_{ij}  \frac{\partial U}{\partial \bm{r}_{j}}
 + k_{B} T B_{ik} \frac{\partial B_{jk}}{\partial \bm{r}_{j}}
 + \sqrt{2 k_{B} T} B_{ij} \bm{w}_{j}^{(\mathrm{S})}(t) ,
\end{equation}
where $B_{ij}(\bm{r}_{1},\bm{r}_{2})$ is the noise coefficient tensor which satisfies
$B_{ik} B_{jk} = L_{ij}$, and $\bm{w}_{i}^{(\mathrm{S})}(t)$ is the Gaussian white noise
which satisfies
\begin{equation}
 \langle \bm{w}_{i}^{(\mathrm{S})}(t) \rangle = 0, \qquad
 \langle \bm{w}_{i}^{(\mathrm{S})}(t) \bm{w}_{j}^{(\mathrm{S})}(t') \rangle = \delta_{ij}\bm{1}\delta(t - t').
\end{equation}
Here, $\bm{1}$ represents the unit tensor.

To study the diffusion and relaxation behavior, it is convenient to use the
bond vector $\bm{q} = \bm{r}_{1} - \bm{r}_{2}$ and the center of mass $\bm{R} = (\bm{r}_{1} + \bm{r}_{2}) / 2$,
instead of the particle positions. The Jacobian determinant is unity for this
variable transform, and thus the potential (the free energy) is not changed.
The bond potential is a function of the bond
vector and is independent of the center of mass, $\partial U / \partial \bm{R} = 0$.
We can rewrite the dynamic equations in terms of $\bm{q}$ and $\bm{R}$.
We calculate the mobility tensor from the dissipative heat flow by utilizing
eq \eqref{mobility_from_heat_flow} (or eq~\eqref{mobility_from_heat_flow_transformed}).
The dissipative heat flow is simply calculated to be
\begin{equation}
 \label{dissipative_heat_flow_bond_cm}
  \dot{\mathcal{Q}}_{d} 
  = - \frac{1}{L_{11}L_{12} - L_{12}^{2}}
 [L_{22} (\dot{\bm{R}} - \dot{\bm{q}} / 2)^{2}
  + L_{11} (\dot{\bm{R}} + \dot{\bm{q}} / 2)^{2}
  - 2 L_{12} (\dot{\bm{R}}^{2} - \dot{\bm{q}}^{2} / 4)] .
\end{equation}
Then, the inverse mobility tensor for $[\bm{q} \  \bm{R}]^{\mathrm{T}}$ is calculated as
\begin{equation}
 \label{inverse_mobility_tensor_bond_cm}
 \begin{split}
  \bm{\Lambda}^{-1} 
  & = - \frac{1}{2}
  \begin{bmatrix}
    {\partial^{2}}/{\partial \dot{\bm{q}} \partial \dot{\bm{q}}} 
   &
   {\partial^{2}}/{\partial \dot{\bm{q}} \partial \dot{\bm{R}}} 
   \\
   {\partial^{2}}/{\partial \dot{\bm{R}} \partial \dot{\bm{q}}}
   &    {\partial^{2}}/{\partial \dot{\bm{R}} \partial \dot{\bm{R}}} 
  \end{bmatrix} 
  \dot{\mathcal{Q}}_{d} \\
  & =
  \frac{1}{L_{11}L_{12} - L_{12}^{2}}
  \begin{bmatrix}
   ( L_{11}  + L_{22}  + 2 L_{12}) / 4
   &
   (L_{11} - L_{22} ) / 2
   \\
   ( L_{11} - L_{22}  ) / 2
   & 
    L_{11}  + L_{22} - 2 L_{12}
  \end{bmatrix} \bm{1}.
 \end{split}
\end{equation}
By inverting eq~\eqref{inverse_mobility_tensor_bond_cm}, we have the mobility tensor as
\begin{equation}
 \label{mobility_tensor_bond_cm}
 \bm{\Lambda}
  =
  \begin{bmatrix}
   L_{11}  + L_{22}  - 2 L_{12}
   &
   (L_{22} - L_{11}) / 2
   \\
   (L_{22} - L_{11}) / 2
   & 
   ( L_{11} + L_{22}  + 2 L_{12}) / 4
  \end{bmatrix} \bm{1}.
\end{equation}
We introduce the noise coefficient tensor for the mobility tensor
\eqref{mobility_tensor_bond_cm} as $\bm{C} \cdot \bm{C}^{\mathrm{T}} = \bm{\Lambda}$.
We use the Cholesky decomposition
of $\bm{\Lambda}$ as the noise coefficient tensor $\bm{C}$:
\begin{equation}
 \bm{C}  = \frac{1}{\sqrt{L_{22} + L_{11} - 2 L_{12}}}
  \begin{bmatrix}
    L_{22} + L_{11} - 2 L_{12}
   & 0 \\
   (L_{22} - L_{11}) / 2
   & \sqrt{L_{11} L_{22}  - L_{12}^{2}}
  \end{bmatrix} \bm{1}.
\end{equation}
With the thus obtained mobility tensor, noise coefficient tensor,
and the bond potential (the free energy), the Langevin equations
for the bond vector and the center of mass can be expressed as
follows:
\begin{equation}
 \label{langevin_equation_bond_cm}
 \frac{d}{dt}
  \begin{bmatrix}
   \bm{q}(t) \\
   \bm{R}(t)
  \end{bmatrix}
  = - \bm{\Lambda}
  \cdot
  \begin{bmatrix}
   \partial U / \partial \bm{q} \\
   0
  \end{bmatrix}
  + k_{B} T 
  \bm{C} \cdot
 \left[ \begin{bmatrix}
	 \displaystyle \frac{\partial}{\partial \bm{q}} &
	 \displaystyle \frac{\partial}{\partial \bm{R}}
 \end{bmatrix}
 \cdot \bm{C} \right]^{\mathrm{T}} 
 + \sqrt{2 k_{B} T} \bm{C} \cdot
 \begin{bmatrix}
  \bm{w}_{1}^{(\mathrm{S})}(t) \\
  \bm{w}_{2}^{(\mathrm{S})}(t)
 \end{bmatrix}.
\end{equation}

{We should recall that} the Langevin equation \eqref{langevin_equation_bond_cm} can
be obtained by directly performing the variable transform for the
Langevin equation. Therefore, although our approach simplified the calculation
somewhat, eq~\eqref{langevin_equation_bond_cm} itself is
rather trivial. As a non-trivial operation, next we consider the coarse-graining.
We eliminate the
degrees of freedom for the bond and derive the effective dynamic
equation for the center of mass. Such a coarse-graining will be useful
when we study the long-time diffusion behavior.
From eq~\eqref{coarse_grained_mobility_tensor}, we have the following expression for the effective mobility:
\begin{equation}
 \label{mobility_tensor_cm}
 \frac{1}{\bar{\Lambda}(\bm{R})} = \frac{1}{\mathcal{Z}} \int d\bm{q} \, \frac{4}{L_{11} + L_{22} + 2 L_{12}} 
  e^{-U(\bm{q}) / k_{B} T}.
\end{equation}
with
$\mathcal{Z} = \int d\bm{q} \, e^{-U(\bm{q}) / k_{B} T}$.
Also, the effective potential (the free energy) $\bar{U}(\bm{R})$ is calculated as
\begin{equation}
 e^{-\bar{U}(\bm{R})} = \int d\bm{q} \, e^{-U(\bm{q}) / k_{B} T},
\end{equation}
and this gives $\bar{U}(\bm{R}) = - k_{B} T \ln \mathcal{Z}$, which is constant of $\bm{R}$.
Therefore the effective potential does not contribute
to any thermodynamic properties, and thus we simply discard it.
The resulting coarse-grained Langevin equation is
\begin{equation}
 \label{langevin_equation_cm}
 \frac{d\bm{R}(t)}{dt} = k_{B} T \sqrt{\bar{\Lambda}(\bm{R})}
  \frac{\partial}{\partial \bm{R}} \sqrt{\bar{\Lambda}(\bm{R})}
  + \sqrt{2 k_{B} T \bar{\Lambda}(\bm{R})} \bm{w}^{(\mathrm{S})}(t),
\end{equation}
where $\bm{w}^{(\mathrm{S})}(t)$ is a Gaussian random noise which satisfies
$\langle \bm{w}^{(\mathrm{S})}(t) \rangle = 0$ and $\langle \bm{w}^{(\mathrm{S})}(t) \bm{w}^{(\mathrm{S})}(t') \rangle
= \bm{1} \delta(t - t')$. Eq~\eqref{langevin_equation_cm} represents a
simple diffusion process with the position-dependent mobility $\bar{\Lambda}(\bm{R})$.
The effect of the eliminated degrees of freedom is taken into account via eq~\eqref{mobility_tensor_cm}.

\subsection{Trapped Particle}

As another simple example, we consider a single particle trapped in a central force potential.
We express the particle position measured from the trap center
as $\bm{r}$ (in the three dimensional Cartesian coordinate), and the trap
potential as $U(\bm{r}) = U(|\bm{r}|)$. The trap potential can be interpreted as the
free energy. We assume that the mobility can depend on the distance from the trap center but
is isotropic, and express it as $L(\bm{r}) = L(|\bm{r}|)$. Then, the Langevin equation
for a particle is
\begin{equation}
  \label{langevin_equation_particle}
  \frac{d  \bm{r}(t)}{d t}
   = - L(\bm{r}) 
    \frac{\partial U(\bm{r})}{\partial {\bm{r}}}
    + k_{B} T \sqrt{L(\bm{r})} \frac{\partial }{\partial \bm{r}} \sqrt{L(\bm{r})}
    + \sqrt{2 k_{B} T L(\bm{r})} \bm{w}^{(\mathrm{S})}(t),
\end{equation}
where $\bm{w}^{(\mathrm{S})}(t)$ is the Gaussian white noise which satisfies
$\langle \bm{w}^{(\mathrm{S})}(t) \rangle = 0$ and
$\langle \bm{w}^{(\mathrm{S})}(t) \bm{w}^{(\mathrm{S})}(t') \rangle = \bm{1} \delta(t - t')$.
The dissipative heat flow is simply calculated as
\begin{equation}
 \label{dissipative_heat_flow_particle}
 \dot{\mathcal{Q}}_{d} = - \frac{\dot{\bm{r}}^{2}}{L(\bm{r})} .
\end{equation}

We consider the dynamic equation in the spherical coordinate {$[ r,\phi,\theta]$ ($r > 0$, $0 \le \phi < 2 \pi$, and $0 \le \theta \le \pi$), defined as
$r = |\bm{r}|$, $\phi = \tan^{-1}(r_{2} / r_{3})$, and
$\theta = \cos^{-1}(r_{3} / r)$}.
The Jacobian determinant is $1 / r^{2} \sin \theta$, and thus the potential
at the spherical coordinate can be expressed as
\begin{equation}
 U'(r,\theta) = U(r)
  - k_{B} T \ln (r^{2} \sin \theta) .
\end{equation}
The dissipative heat flow \eqref{dissipative_heat_flow_particle} can be rewritten in terms of $[\dot{r}, \dot{\phi}, \dot{\theta}]$ as
\begin{equation}
 \dot{\mathcal{Q}}_{d} = - \frac{1}{L(r)}
  \left[ \dot{r}^{2} 
   + r^{2} \sin^{2} \theta \, \dot{\phi}^{2}
  + r^{2} \dot{\theta}^{2}  \right] ,
\end{equation}
and thus the inverse mobility tensor for ${[r,\phi,\theta]^{\mathrm{T}}}$ becomes
\begin{equation}
 \label{inverse_mobility_tensor_particle}
 {\bm{L}'}^{-1} = \frac{1}{L(r)}
  \begin{bmatrix}
   1 & 0 & 0 \\
   0 & r^{2} \sin^{2} \theta & 0 \\
   0 & 0 & r^{2} 
  \end{bmatrix} .
\end{equation}
By inverting eq~\eqref{inverse_mobility_tensor_particle}, we have
the mobility tensor and the noise coefficient tensor as
\begin{equation}
 \label{mobility_tensor_particle}
 \bm{L}' = L(r)
  \begin{bmatrix}
   1 & 0 & 0 \\
   0 & 1 / r^{2} \sin^{2} \theta & 0 \\
   0 & 0 & 1 / r^{2}
  \end{bmatrix}, \qquad
 \bm{B}' = \sqrt{L(r)}
  \begin{bmatrix}
   1 & 0 & 0 \\
   0 & 1 / r \sin \theta & 0 \\
   0 & 0 & 1 / r
  \end{bmatrix} .
\end{equation}
Therefore, the Langevin equations for $r,\phi,$ and $\theta$ become
\begin{equation}
 \label{langevin_equation_particle_spherical_r}
 \frac{dr(t)}{dt}
  =  - L(r) \left[ \frac{d U(r)}{d r}
    - \frac{2 k_{B} T}{r}
   \right]
  + \frac{k_{B} T}{2}  \frac{d L(r)}{d r}
  + \sqrt{2 k _{B} T L(r)} w_{1}^{(\mathrm{S})}(t), 
\end{equation}
\begin{equation}
 \label{langevin_equation_particle_spherical_theta}
 \frac{d\phi(t)}{dt}
  =  \frac{\sqrt{2 k _{B} T L(r)}}{r \sin \theta} w_{2}^{(\mathrm{S})}(t) ,  
\end{equation}
\begin{equation}
 \label{langevin_equation_particle_spherical_phi}
 \frac{d\theta(t)}{dt}
  = - \frac{L(r)}{r^{2}} 
  \left( - \frac{ k_{B} T }{\tan \theta} \right)
  + \frac{\sqrt{2 k_{B} T L(r)}}{r} w_{3}^{(\mathrm{S})}(t).
\end{equation}
The potential $U'(r,\theta)$ gives an additional force term
for $r$
arising from the Jacobian determinant (Neumann's radial force\cite{Neumann-1980}), as
clearly found in eq~\eqref{langevin_equation_particle_spherical_r}.
This additional force diverges at $r \to 0$, and thus it repels the particle from
the trap center. We observe a similar additional force term for $\theta$ in
eq~\eqref{langevin_equation_particle_spherical_phi}.
The derivation of eqs~\eqref{langevin_equation_particle_spherical_r}-\eqref{langevin_equation_particle_spherical_theta}
based on our procedure is straightforward, and free from many
complicated and lengthy calculations for derivatives in spherical coordinates.

Next, we consider to derive an approximate dynamic equation for the 
distance from the trap center, $r$. This can be achieved {by} the coarse-graining,
in which the remaining degrees of freedom, $\phi$ and $\theta$, are eliminated.
The effective potential
for $r$, $\bar{U}(r)$, is calculated as
\begin{equation}
 e^{-\bar{U}(r) / k_{B} T} = \int d\phi d\theta \, r^{2} \sin \theta \, e^{-U(r) / k_{B} T}
  = 4 \pi r^{2} e^{-U(r) / k_{B} T},
\end{equation}
and we have
$\bar{U}(r) = U(r) - k_{B} T \ln r^{2} + \text{(const.)}$.
The terms in the dissipative heat flow \eqref{dissipative_heat_flow_particle} are
already decoupled, thus we simply have the effective mobility $\bar{L}(r)$ for $r$ as
\begin{equation}
 \label{inverse_effective_mobility_particle}
  \frac{1}{\bar{L}(r)} = \int d\phi d\theta \, \frac{\sin \theta}{4 \pi} \frac{1}{L(r)} = \frac{1}{L(r)} .
\end{equation}
Finally, we have the following Langevin equation as an approximate dynamic
equation for $r$:
\begin{equation}
  \label{langevin_equation_particle_r}
 \frac{dr(t)}{dt}
  = - L(r) 
  \left[ \frac{d U(r)}{d r}
    - \frac{2 k_{B} T}{r}
   \right]
  + \frac{k_{B} T}{2} \frac{d L(r)}{d r} 
  + \sqrt{2 k_{B} T L(r)} w^{(\mathrm{S})}(r),
\end{equation}
where $w^{(\mathrm{S})}(t)$ is the Gaussian white noise which satisfies $\langle w^{(\mathrm{S})}(t) \rangle = 0$ and
$\langle w^{(\mathrm{S})}(t) w^{(\mathrm{S})}(t') \rangle = \delta(t - t')$.
Eq~\eqref{langevin_equation_particle_r} has the same form as
eq~\eqref{langevin_equation_particle_spherical_r}. Because the dynamics of $\theta$ and $\phi$
does not affect that of $r$, this result is reasonable.

We consider other approximate dynamic equations, by eliminating the
distance $r$. This is another coarse-graining and the effective directional
dynamics will be obtained. The effective potential for $\phi$ and $\theta$ is
\begin{equation}
 e^{-\bar{U}(\phi,\theta)} = \sin \theta \int dr \, r^{2} e^{-U(r) / k_{B} T},
\end{equation}
and thus $\bar{U}(\phi,\theta) = -k_{B} T \ln \sin \theta + (\text{const}.)$.
(The same effective potential was utilized by Doi~\cite{Doi-book},
although he showed the free energy for the distribution which also has the entropic contribution.)
The effective mobility tensor $\bar{\bm{L}}$ for ${[\phi,\theta]^{\mathrm{T}}}$ is given as
\begin{equation}
 \bar{\bm{L}}^{-1}
  = \frac{1}
  {\displaystyle \int dr \, r^{2} e^{-U(r) / k_{B} T}}
 \int dr \, r^{2} e^{-U(r) / k_{B} T} \frac{1}{L(r)}
  \begin{bmatrix}
   r^{2} \sin^{2} \theta & 0 \\
   0 & r^{2}
  \end{bmatrix}.
\end{equation}
Although we cannot calculate the integrals over $r$ unless the potential
$U(r)$ is specified, the $r$-dependent part can be separated and we have
the following general form:
\begin{equation}
 \bar{\bm{L}}
  = \bar{\Lambda}
  \begin{bmatrix}
   1 /  \sin^{2} \theta & 0 \\
   0 & 1
  \end{bmatrix} ,
\end{equation}
with
\begin{equation}
  \bar{\Lambda}
  = \frac{\displaystyle \int dr \, (r^{4} / L(r)) e^{-U(r) / k_{B} T} }
  {\displaystyle \int r^{2} e^{-U(r) / k_{B} T}} .
\end{equation}
The effective dynamic equations for $\phi$ and $\theta$ can be expressed as
follows:
\begin{equation}
 \label{langevin_equation_particle_direction_theta}
 \frac{d\phi(t)}{dt}
  =  \frac{\sqrt{2 k _{B} T \bar{\Lambda}}}{\sin \theta} w_{2}^{(\mathrm{S})}(t) ,  
\end{equation}
\begin{equation}
 \label{langevin_equation_particle_direction_phi}
 \frac{d\theta(t)}{dt}
  = - \bar{\Lambda}
  \left( - \frac{ k_{B} T }{\tan \theta} \right)
  + \sqrt{2 k_{B} T \bar{\Lambda}} w_{3}^{(\mathrm{S})}(t).
\end{equation}
Unlike the case of the effective dynamic equation for $r$, 
eqs~\eqref{langevin_equation_particle_direction_theta}
and \eqref{langevin_equation_particle_direction_phi} are different from
eqs~\eqref{langevin_equation_particle_spherical_theta}
and \eqref{langevin_equation_particle_spherical_phi}.
By performing the coarse-graining,
$L(r) / r^{2}$ in eqs~\eqref{langevin_equation_particle_spherical_theta}
and \eqref{langevin_equation_particle_spherical_phi} is replaced by $\bar{\Lambda}$.
The factor $\bar{\Lambda}$ depends on the mobility $L(r)$ and the bond potential $U(r)$,
and the characteristic
relaxation time  reflects the information on $L(r)$ and $U(r)$.

\subsection{Density Field of Brownian Particles}

As the last example, we consider the overdamped interacting particles.
We describe the position of the $i$-th particle as $\bm{R}_{i}$.
We assume that the particles interact via the pairwise interaction potential $u(\bm{r})$, and the 
interaction energy is given as $U(\lbrace \bm{R}_{i} \rbrace) = \sum_{i > j} u(\bm{R}_{i} - \bm{R}_{j})$.
The Langevin equation for $\bm{R}_{i}$ is
\begin{equation}
 \label{langevin_equation_interacting_particles}
 \frac{d\bm{R}_{i}(t)}{dt} = - \frac{\partial U(\lbrace \bm{R}_{i} \rbrace) }{\partial \bm{R}_{i}}
  + \sqrt{\frac{2k_{B}T}{\zeta}}\bm{w}_{i}^{(\mathrm{S})}(t),
\end{equation}
where $\zeta$ is the friction coefficient, and $\bm{w}^{(\mathrm{S})}_{i}(t)$ is the Gaussian white noise which satisfies the
fluctuation-dissipation relation:
\begin{equation}
 \langle \bm{w}_{i}^{(\mathrm{S})}(t) \rangle = 0, \qquad
 \langle \bm{w}_{i}^{(\mathrm{S})}(t) \bm{w}_{j}^{(\mathrm{S})}(t') \rangle = \delta_{ij} \bm{1} \delta(t - t').
\end{equation}

To study the dynamic of this system, the dynamic equation for the density
field is sometimes useful. The (microscopic) density field is defined as
\begin{equation}
 \label{microscopic_density_field_definition}
 \hat{\rho}(\bm{r},t) \equiv \sum_{i} \delta(\bm{r} - \bm{R}_{i}(t)) .
\end{equation}
Dean\cite{Dean-1996} derived the dynamic equation for the density field starting from
the Langevin equation \eqref{langevin_equation_interacting_particles},
by using the Ito calculus. Also, we can calculate the mobility tensor for
the density field by the transform in eq~\eqref{mobility_tensor_transformed}.
Here we construct the dynamic equation in another way. The dissipative heat
flow is calculated from eq \eqref{langevin_equation_interacting_particles} as
\begin{equation}
 \label{dissipative_heat_flow_interacting_particles}
  \dot{\mathcal{Q}}_{d}
  \left(\lbrace \bm{R}_{i} \rbrace,\left\lbrace \frac{d\bm{R}_{i}}{dt} \right\rbrace\right)
 = - \sum_{i} \zeta \left( \frac{d\bm{R}_{i}(t)}{dt} \right)^{2} .
\end{equation}
From eq \eqref{microscopic_density_field_definition}, we have
\begin{equation}
 \label{microscopic_density_field_time_derivative}
 \frac{\partial \hat{\rho}(\bm{r},t)}{\partial t} =
 J_{i\alpha}(\bm{r}) \frac{dR_{i\alpha}(t)}{dt} ,
\end{equation}
where $R_{i\alpha}$ represents the $\alpha$-th component of the vector $\bm{R}_{i}$, 
and the summation is taken for $i$ and $\alpha$. (The Einstein convention
is applied.)
We have defined 
\begin{equation}
 \bm{J}_{i}(\bm{r}) \equiv 
  - \frac{\partial}{\partial \bm{r}} \delta(\bm{r} - \bm{R}_{i}(t)) ,
\end{equation}
and interpret this $J_{i\alpha}(\bm{r})$ is a rectangular matrix
(the dimensions for $i\alpha$ and $\bm{r}$ are different).
We introduce the Moore-Penrose pseudo inverse\cite{BenIsrael-Greville-book,Petersen-Pedersen-book} $J_{i\alpha}^{+}(\bm{r})$ via:
$ \int d\bm{r}' J_{j\beta}(\bm{r}) J_{j\beta}^{+}(\bm{r}') J_{i\alpha}(\bm{r}')
  = J_{i\alpha}(\bm{r}) $.
Then eq \eqref{microscopic_density_field_time_derivative} can be inverted
as
\begin{equation}
 \label{microscopic_density_field_time_derivative_modified}
  \frac{dR_{i\alpha}(t)}{dt} =
  \int d\bm{r} \, J_{i\alpha}^{+}(\bm{r},t) \frac{\partial \hat{\rho}(\bm{r},t)}{\partial t} .
\end{equation}
Here it would be fair to mention that eq \eqref{microscopic_density_field_time_derivative_modified} is
not a unique solution of eq \eqref{microscopic_density_field_time_derivative}.
The Moore-Penrose pseudo inverse gives the most reasonable solution and there can be other solutions.
Here we simply
employ eq \eqref{microscopic_density_field_time_derivative_modified} as the solution.
From eqs \eqref{dissipative_heat_flow_interacting_particles} and \eqref{microscopic_density_field_time_derivative_modified},
the dissipative heat flow can be rewritten as
\begin{equation}
 \label{dissipative_heat_flow_interacting_particles_modified}
  \dot{\mathcal{Q}}_{d}
  \left[ \hat{\rho}, \frac{\partial \hat{\rho}}{\partial t} \right] = 
  - \int d\bm{r} d\bm{r}' \, \zeta J_{i\alpha}^{+}(\bm{r})
  J_{i\alpha}^{+}(\bm{r}')  \frac{\partial \hat{\rho}(\bm{r},t)}{\partial t} 
   \frac{\partial \hat{\rho}(\bm{r}',t)}{\partial t} .
\end{equation}

From eq \eqref{dissipative_heat_flow_interacting_particles_modified}, the
inverse mobility for the density field becomes
\begin{equation}
 \label{inverse_mobility_tensor_interacting_particles}
  L^{-1}(\bm{r},\bm{r}') = \zeta 
   J_{i\alpha}^{+}(\bm{r})
  J_{i\alpha}^{+}(\bm{r}') 
  = \zeta  [J_{i\alpha}(\bm{r})
  J_{i\alpha}(\bm{r}')]^{-1}.
\end{equation}
Here we utilize the explicit expression for the Moore-Penrose pseudo inverse: $J_{i\alpha}^{+}(\bm{r}) 
= \int d\bm{r}' \, J_{i\alpha}(\bm{r}') $ $[J_{j\beta}(\bm{r}') J_{j\beta}(\bm{r})]^{-1} $.
Therefore, we have the following simple expression for the mobility tensor:
\begin{equation}
 \label{mobility_for_density_field_final}
\begin{split}
 L(\bm{r},\bm{r}')
 & = \frac{1}{\zeta}  J_{i\alpha}(\bm{r})
  J_{i\alpha}(\bm{r}') 
 = \frac{1}{\zeta}  \sum_{i} \frac{\partial}{\partial \bm{r}}\delta(\bm{r} - \bm{R}_{i}) 
 \cdot \frac{\partial}{\partial \bm{r}'} \delta(\bm{r}' - \bm{R}_{i}) 
 \\
 & = \frac{\partial}{\partial \bm{r}} \cdot 
  \left[ \frac{\hat{\rho}(\bm{r})}{\zeta} \frac{\partial}{\partial \bm{r}'} 
  \delta(\bm{r} - \bm{r}') \right].
\end{split}
\end{equation}
From our derivation, eq~\eqref{mobility_for_density_field_final} may not be exact, but it works as
the most reasonable expression, which would be a good approximation in many cases.
(Fortunately, the exact derivation by Dean\cite{Dean-1996} gives the same mobility tensor and
thus our method works perfectly for this system.)
By combining the mobility tensor \eqref{mobility_for_density_field_final}
with the free energy for the density field\cite{Hansen-McDonald-book,Dean-1996}
\begin{equation}
 \label{free_energy_for_density_field}
 \mathcal{F}[\hat{\rho}]
  =  k_{B} T \int d\bm{r} \,\hat{\rho}(\bm{r}) [ \ln \hat{\rho}(\bm{r}) - 1 ]
  + \frac{1}{2} \int d\bm{r} d\bm{r}' \hat{\rho}(\bm{r}) u(\bm{r} - \bm{r}') \hat{\rho}(\bm{r}') ,
\end{equation}
we can construct the Langevin equation for the density field. For convenience, we introduce the
following noise coefficient tensor
\begin{equation}
 \bm{B}(\bm{r},\bm{r}') = - \sqrt{\frac{\hat{\rho}(\bm{r}')}{\zeta}}  \frac{\partial}{\partial \bm{r}'} \delta(\bm{r} - \bm{r}'),
\end{equation}
which satisfies $\int d\bm{r}' \bm{B}(\bm{r},\bm{r}') \cdot \bm{B}(\bm{r}'',\bm{r}') = L(\bm{r},\bm{r}'')$.
Then the Ito type Langevin equation becomes as follows, which is equivalent to the Dean's equation\cite{Dean-1996}:
\begin{equation}
 \label{dean_equation}
\begin{split}
   \frac{\partial \hat{\rho}(\bm{r},t)}{\partial t}
 & = \int d\bm{r}' \, \left[ - L(\bm{r},\bm{r}')  \frac{\delta \mathcal{F}[\hat{\rho}]}{\delta \hat{\rho}(\bm{r}',t)} 
 + k_{B} T \frac{\delta L(\bm{r},\bm{r}') }{\delta \hat{\rho}(\bm{r}',t)}
  + \sqrt{2k_{B}T} \bm{B}(\bm{r},\bm{r}') \cdot \bm{w}^{(\mathrm{I})}(\bm{r}',t) \right] \\
 & = \frac{\partial}{\partial \bm{r}} \cdot 
  \left[ \frac{\hat{\rho}(\bm{r},t)}{\zeta}
 \frac{\partial}{\partial \bm{r}} \frac{\delta \mathcal{F}[\hat{\rho}]}{\delta \hat{\rho}(\bm{r},t)} \right]
  + \frac{\partial}{\partial \bm{r}} \cdot \left[ \sqrt{\frac{2 k_{B} T \hat{\rho}(\bm{r},t)}{\zeta}} \bm{w}^{(\mathrm{I})}(\bm{r},t) \right],
\end{split}
\end{equation}
where $\bm{w}^{(\mathrm{I})}(\bm{r},t)$ is the Gaussian white noise field
which satisfies
\begin{equation}
 \langle \bm{w}^{(\mathrm{I})}(\bm{r},t) \rangle = 0, \qquad
 \langle \bm{w}^{(\mathrm{I})}(\bm{r},t) \bm{w}^{(\mathrm{I})}(\bm{r}',t') \rangle = \bm{1} \delta(\bm{r} - \bm{r}') \delta(t - t').
\end{equation}
If we ignore the stochastic term in eq \eqref{dean_equation},
it reduces to the diffusion type equation under the mean-field approximation.

As Dean showed, eq \eqref{dean_equation} can be directly obtained from
the Langevin equation \eqref{langevin_equation_interacting_particles}
{in as straightforward way}. The Langevin equation for the density
field itself is not a new result. Moreover, the original derivation by Dean
is {simple} and does not need to handle the mobility tensor explicitly.
Nonetheless our method will be useful
because it provides an alternative way to construct the Langevin equation.
The interpretation of the Dean's equation is not that simple, and
various {studies} have been done\cite{Marconi-Tarazona-1999,Frusawa-Hayakawa-2000,Archer-Rauscher-2004,DuranOlivencia-Yatsyshin-Goddard-Kalliadasis-2018}.
In the Dean's original derivation, the Langevin equation for the
density field is directly derived from the Langevin equation for particles
(eq \eqref{langevin_equation_interacting_particles}).
The derivation itself is mathematically straightforward, but the {physical} interpretation
of the microscopic density field and how {to} perform the coarse-graining
is not that clear\cite{Archer-Rauscher-2004,DuranOlivencia-Yatsyshin-Goddard-Kalliadasis-2018}.
In the thus obtained dynamic
equation, the mobility tensor and the free energy are mixed up and one should
separate them manually. On the other hand, our method directly gives
the expression of the
mobility tensor (although the calculation is not simple).
The free energy is obtained
from the partition function, and this calculation is independent of the
Langevin equation. Also, our method can be applied for the coarse-graining
(at least in principle). We expect that our method will provide a new aspect when we study
the Dean's and related dynamic equations.

Doi\cite{Doi-book} claimed that a similar diffusion type
equation can be obtained by the Onsager's method. However, in the Doi's derivation,
the Rayleigh dissipation function is expressed in terms of the particle velocity field $\hat{\bm{v}}(\bm{r},t)$.
The velocity field is defined via 
\begin{equation}
 \label{microscopic_velocity_field_via_conservation_law}
 \frac{\partial \hat{\rho}(\bm{r},t)}{\partial t}
  = - \frac{\partial}{\partial \bm{r}} \cdot [ \hat{\rho}(\bm{r},t) \hat{\bm{v}}(\bm{r},t) ],
\end{equation}
and the Rayleigh's dissipation function is assumed to be
\begin{equation}
 \label{dissipation_function_empirical}
 \Psi = \frac{1}{2} \int d\bm{r} \, \zeta \hat{\rho}(\bm{r}) \hat{\bm{v}}^{2}(\bm{r}).
\end{equation}
Then, Doi constructed the Rayleighian from eq~\eqref{dissipation_function_empirical} and the free energy
\eqref{free_energy_for_density_field} and derived
the dynamic equation which is similar to eq~\eqref{dean_equation}
(without the noise term) by minimizing the Rayleighian {\em with respect to $\hat{\bm{v}}(\bm{r})$}.
However, this derivation lacks the consistency. The Rayleigh's dissipation function (which
is essentially the same as the dissipative heat, from eq~\eqref{dissipative_heat_flow_and_rayleigh_dissipation_function})
should be interpreted as a functional of the slow variable $\hat{\rho}(\bm{r})$ and
{\em its time derivative} $\partial \hat{\rho}(\bm{r}) / \partial t$. To be consistent, the minimization should
be performed for $\partial \hat{\rho}(\bm{r}) / \partial t$, not for other variables such as $\hat{\bm{v}}(\bm{r})$.
(Doi\cite{Doi-Zhou-Di-Xu-2019} also showed a similar minimization with respect to another variable $\hat{\bm{j}}(\bm{r}) \equiv \hat{\rho}(\bm{r}) \hat{\bm{v}}(\bm{r})$.)
On the other hand, in our derivation, the dissipation function {(the dissipative heat flow)} is interpreted as a functional of
$\hat{\rho}(\bm{r})$ and  $\partial \hat{\rho}(\bm{r}) / \partial t$, and
thus is systematic and fully consistent with the general method.
(Doi's derivation gives the correct dynamic equation because
the variable $\hat{\bm{v}}(\bm{r})$ is linear in $\partial \hat{\rho}(\bm{r}) / \partial t$.
If the variable for the minimization is not linear in $\partial \hat{\rho}(\bm{r}) / \partial t$, of course,
the resulting dynamic equation becomes incorrect.)

\section{Conclusions}

We analyzed the dissipation of the mesoscopic systems which obey the Langevin equation,
from the view point of the stochastic energetics. From the energy balance
equation, we obtained the dissipative and fluctuating heat flows. The
dissipative heat flow has a similar form to the Rayleigh dissipation
function, but the physical meaning is much {clearer}; the dissipative
heat flow simply represents the heat which moves from the bath to the system
in absence of the fluctuation.

We showed that the dissipative heat flow can be used to calculate
the (inverse) mobility. The dissipative heat flow can be related to the
mobility tensor, and in fact, we can systematically calculate the mobility tensor
once the dissipative heat flow is obtained.
We also showed that the dissipative heat flow
is in a covariant form, whereas the free energy is not in a covariant form.
(This means that, neither our energy balance equation nor
the Rayleighian is covariant.)
To construct the Langevin equation for
a specific variable, we can utilize the dissipative heat flow calculated
for other variables. Moreover, the dissipative heat flow can be utilized
to perform the coarse-graining. The mobility tensor of the coarse-grained
variables can be systematically constructed by calculating the approximate
dissipative heat flow in local equilibrium.
Then we can construct the Langevin equation by combining the free energy
and the obtained mobility tensor.
Our method is similar to but qualitatively different
from the Reyleigh's and Onsager's methods, and can be utilized as an
alternative to them.

As examples, we applied our method to several systems such as the
dumbbell model and the Brownian particles. Our method can successfully reproduce
the Dean's equation for the density field of the interacting Brownian particles.
Our method gives
the dynamic equations and coarse-grained effective dynamic equations for these
systems which are consistent with those obtained by conventional method.
We consider that our method will be useful to construct
the dynamic equation models to study rheological properties of some soft matter systems.
Of course, the usefulness strongly depends on {the properties of} the target system.
Judging from the results in this work,
we conclude that there is no general construction method for mesoscopic and macroscopic dynamic equations,
which is variational and covariant.
Our method may be convenient for some cases, yet
the conventional methods without the dissipative heat flow (and the Rayleigh dissipation function) may be
more convenient for some other cases.
We should emphasize that our method (or any other method) is never a ``silver bullet.''
Our method is just one candidate
among various methods, and one should carefully choose an appropriate
method to construct dynamic equations.

\section*{Acknowledgment}

The author thanks Prof.~M.~Doi (Beihang University) for comments and discussions.
This work was supported by JST, PRESTO Grant Number JPMJPR1992, Japan,
and Grant-in-Aid (KAKENHI) for Scientific Research B JP19H01861.

\end{document}